\documentclass[twocolumn,10pt]{autart}    
\usepackage{graphicx}          
\usepackage{amsmath}
\usepackage{enumerate}
\usepackage{graphicx}          
\usepackage{tikz}
\usepackage{cite}
\usepackage{natbib}

\usepackage{amsfonts}

\usepackage{amssymb}
\usepackage{subfigure}
\usepackage{hyperref}
\hypersetup{
	colorlinks = true,
	citecolor = cyan,
}
\usepackage{wrapfig}

\newtheorem{remark}{Remark}

\newtheorem{assumption}{Assumption}
\newtheorem{definition}{Definition}
\newtheorem{lemma}{Lemma}
\newtheorem{corollaryx}{Corollary}
\newtheorem{theorem}{Theorem}

\begin{document}
	\begin{frontmatter}
		
		\title{Identification and Adaptation with Binary-Valued Observations under Non-Persistent Excitation Condition\thanksref{footnoteinfo}} 
		\thanks[footnoteinfo]{This work was supported by the National Natural Science Foundation of China under Grants No. 11688101 and 62025306.}
		\thanks[footnote]{Corresponding author: Lei Guo.}
		\author[AMSS,UCAS]{Lantian Zhang}\ead{zhanglantian@amss.ac.cn},    
		\author[AMSS,UCAS]{Yanlong Zhao}\ead{ylzhao@amss.ac.cn},               
		\author[AMSS,UCAS]{Lei Guo\thanksref{footnote}}\ead{lguo@amss.ac.cn}  
		
		\address[AMSS]{Key Laboratory of Systems and Control, Academy of Mathematics and Systems Science, Chinese Academy of Sciences, Beijing 100190, China}
		\address[UCAS]{School of Mathematical Science, University of Chinese Academy of Sciences, Beijing 100049, China}  


		\begin{abstract}                          
Dynamical systems with binary-valued observations are widely used in information industry, technology of biological pharmacy and other fields. Though there have been much efforts devoted to the identification of such systems, most of the previous investigations are based on first-order gradient algorithm which usually has much slower convergence rate than the Quasi-Newton algorithm. Moreover, persistence of excitation(PE) conditions are usually required to guarantee consistent parameter estimates in the existing literature, which are hard to be verified or guaranteed for feedback control systems. In this paper, we propose an online projected Quasi-Newton type algorithm for parameter estimation of stochastic regression models with binary-valued observations and varying thresholds.  By using both the stochastic Lyapunov function and martingale estimation methods, we establish the  strong consistency of the estimation algorithm and provide the convergence rate, under a signal condition which is considerably weaker than the traditional PE condition and coincides with the weakest possible excitation known for the classical least square algorithm of stochastic regression models. Convergence of adaptive predictors and their applications in adaptive control are also discussed.
		\end{abstract}
	\end{frontmatter}
	
	\section{Introduction}
	\subsection{Background}
	The purpose of this paper is to study parameter estimation and adaptation for stochastic systems, in which the system output cannot be measured accurately, and the only available information is whether or not the output belongs to some set (\cite{zhang:2003}). Specifically,  consider the following standard stochastic regression model:
	\begin{equation}\label{1}
		y_{k+1}=\phi_{k}^{T} \theta+v_{k+1}, \quad k=0, 1,2, \ldots
	\end{equation}
	where $y_{k+1}\in R^{1}, v_{k+1}\in R^{1}, \phi_{k}\in R^{p}(p\geq1)$ represent the system output, random noise and regression vector, respectively, and  $\theta \in R^{p}$ is an unknown parameter vector to be estimated. The system output $y_{k+1}$  can only be observed with binary-valued measurements:
	
	\begin{equation}\label{2}
		s_{k+1}=I\left(y_{k+1} \geq c_{k}\right)=\left\{\begin{array}{ll}
			1, &  y_{k+1} \geq c_{k};\\
			0, & otherwise,
		\end{array}\right.
	\end{equation}	
	where $\{c_{k}\}$ denotes a given threshold sequence, and $I(\cdot)$ is the indicator function.
		
	This type of observations emerges widely in practical systems with the development of science and technology. One example comes from neuron systems(\cite{ghysen:2003}) where, instead of measuring exact internal potential, the systems only provide information of states (excitation or inhibition). When the potential is smaller than the potential threshold, the neuron shows the inhibition state, otherwise shows the excitation state. The objective of classification via neural network is to learn the system parameters based on states of neurons only. Another example comes from sensor networks(\cite{zhang:2019}) where, for set-valued sensor networks, the information from each sensor turns out to be quantized observations with finite number of bits or even 1 bit. Specifically, each sensor only provide information whether the measured value is larger than a designed threshold or not. Usually, such sensors are more cost effective than regular sensors. Besides, there are also numerous other examples such as ATM ABR traffic control;  gas content sensors ($CO_{2}$, $H_{2}$, etc.) in gas and oil industry and switching sensors for shift-by-wire in automotive applications(\cite{zhang:2003}).
	
	\subsection{Related works}
	Due to the widely use of systems with quantized observations, a number of basic problems concerning identification and control  emerge, which need theoretical investigations. In fact, the estimation of quantized output systems has constituted a vast literature recently. \cite{zhang:2003}; \cite{yin:2007}; \cite{yin:2008}; \cite{yin:2010} gave a strongly consistent identification algorithm under periodic signals.  \cite{jafari:2012} proposed a recursive identification algorithm for FIR systems with binary-valued observations, and proved its convergence under $\alpha$-mixing property of the signals. Later, \cite{guo:2013} proposed a recursive projection algorithm for FIR systems with binary-valued observations and fixed thresholds. The paper established the convergence rate $O\left(\frac{\log k}{k}\right)$ under the following strongly persistent excitation condition $\sum_{l=k}^{k+N-1} \phi_{l} \phi_{l}^{T} \geq \varepsilon I, \quad k=1,2, \ldots$, for some fixed $N$ and constant $\epsilon >0$. Besides, \cite{you:2013} considered ARMA system with time-varying K -
level scalar quantizer also  random packet dropouts,  and gave the consistence result under independent and identically distributed(i.i.d.) conditions. The adaptive quantizer is considered in \cite{you:2015} for FIR systems, provided that the signals satisfy i.i.d conditions with $E[\phi_{i}\phi_{i}^{\tau}]=H >0$.  Moreover,  under some persistent excitation conditions, \cite{zhao:2016}  introduced a EM-typed algorithm, which is robust and easy to programme, and proved that the maximum likelihood criterion can be achieved as the number of iterations goes to infinity.  \cite{song:2018} presented a strongly consistent estimate and obtained the convergence rate for ARMA systems with binary sensors and unknown threshold under i.i.d Gaussian inputs.  \cite{zhang:2019} considered a Quasi-Newton type algorithm under the following persistent excitation (PE) condition: $\liminf _{k \rightarrow \infty} \lambda_{min}\left\{\frac{1}{k} \sum_{i=1}^{k} \phi_{i} \phi_{i}^{T}\right\}>0,$ where $\lambda_{min}\left\{\cdot \right\}$ represents the minimal eigenvalue of the matrix in question. Numerical simulations in \cite{zhang:2019} demonstrated that their Quasi-Newton type algorithm has equivalent convergence properties for first-order FIR systems and high-order systems, where strong consistency and  asymptotic efficiency for first-order FIR systems are also established. For other methods, We refer the readers to \cite{bottegal:2017},  \cite{wang:2014} about kernel-based method and quadratic programming-based method, among others.
	
	However, almost all of the existing investigations on identification suffer from some fundamental limitations.
	
	Firstly, for the system with regular output sensors, i.e.,  $s_{k+1}=y_{k+1}$ in $(\ref{2})$, substantial progresses had been made in  the area of adaptive estimation and control (e.g.,\cite{chen:1991}), and the excitation condition for consistency of parameter estimates need not to be persistent. For example, it is widely known that the weakest possible excitation condition for strong consistency of the least-squares estimate for stochastic regression models is the following (\cite{lai:1982}):
	\begin{equation}\label{3}
		\left\{\log \lambda_{\max }\left(\sum_{i=1}^{n} \phi_{i} \phi_{i}^{T}\right)\right\} / \lambda_{\min }\left(\sum_{1}^{n} \phi_{i} \phi_{i}^{T}\right) \rightarrow 0. \quad a.s.,
	\end{equation}
	which is actually a decaying excitation condition, is much weaker than the classical PE condition, and can be applied in adaptive feedback control(see e.g., \cite{chen:1991}). However, as mentioned above, for identification of systems with binary-valued sensors, almost all of the existing literatures need the PE conditions on signals for strong consistency, and actually, most need i.i.d or periodic signal assumptions. Though these conditions may be satisfied for some open-loop or off-line identification, they are  much more difficult to be satisfied or verified for closed-loop system identification, since the input and output data of such systems are generally determined by nonlinear stochastic dynamic equations(\cite{chen:1991}). Consequently, the problem whether or not the PE condition can be essentially weakened to e.g. (\ref{3}) for identification of stochastic systems with binary-valued sensors still remains open.
	
	Secondly, to the best of the authors' knowledge, almost all of the existing estimation algorithms for stochastic systems with binary-valued observations and given thresholds, are designed with first-order gradient. This kind of algorithms is designed by taking the same step-size for each coordinates, which may alleviate the complexity in the convergence analysis, but will sacrifice the convergence rate of the algorithms (\cite{ljung:1983}). To improve the convergence properties, it is necessary to consider estimation algorithms with adaptation gain being a matrix (e.g., Hessian matrix or its modifications), rather than a scalar.
	
	Thirdly, there are only a few results on adaptive control with binary-valued observations in the existing literature (c.f.,e.g., \cite{guo:2011}, \cite{zhao:2013}), where some kinds of FIR control systems are considered and consistency of parameter estimates is needed for the optimality of adaptive control systems.

    The goal of this paper is to show that the above mentioned limitations can be considerably relaxed or removed.
	
	\subsection{Contributions}
	Inspired by the method in \cite{guo:1995}, this paper proposes a new recursive projected Quasi-Newton type algorithm which can be viewed as a naturally extension of classical linear least-square algorithm with a projection operator. The main contributions of this paper can be summarized as follows:
	\begin{itemize}
		\item We propose a projected recursive Quasi-Newton type algorithm for stochastic regression systems with binary-valued observations. In the area of  identification with binary-valued observations and given fixed thresholds, this paper appears to be the first to establish almost sure convergence for Quasi-Newton type estimation algorithms where the adaptation gains are matrices. \\
		
		\item The weakest possible excitation condition $(\ref{3})$ known for strong consistency of the classical least-squares algorithm, is proven to be sufficient for strong consistency of the proposed new estimation algorithm in the current binary-valued observation case.  This appears to be the first time to achieve such a strong result in the literature of system identification with binary-valued observations. \\
				
		\item We also obtain a celebrated result on the asymptotic order of the accumulated regret of adaptive prediction, that is $\sum_{k=0}^{n}\left[E\left(y_{k+1} \mid \mathcal{F}_{k}\right)-\hat{y}_{k+1}\right]^{2}=O(\log n)$,  $a.s$, which does not need any excitation condition and can be conveniently used in adaptive control to give better results than the existing ones in the literature.
	\end{itemize}

	The remainder of this paper is organized as follows. In Section 2, we give the main results of this paper, including the assumptions, proposed algorithms and main theorems; Section 3 presents the proofs of the main results together with some key lemmas. Some numerical examples are provided in Section 4. Finally, we conclude the paper with some remarks.
	
	\section{The main results}
	
	Consider the stochastic regression model (\ref{1})-(\ref{2}) with binary-valued observations. The objectives of this paper are, to propose a strongly consistent estimator $\{\hat{\theta}_{n}\}$ for the unknown parameter vector $\theta$ under a non-PE condition, and to give an asymptotically optimal adaptive predictor $\{\hat{y}_{n+1}\}$ for the regular output $\{y_{n+1}\}$  together with its applications in adaptive tracking.
	\subsection{Notations and assumptions}
	For our purpose, we introduce some notations and assumptions first.
	
	{\bf Notations.}  By $\|\cdot\|$, we denote the Euclidean-norm of vectors or matrices. The spectrum of a symmetric matrix $M$ is denoted by $\left\{\lambda_{i}\left\{M\right\}\right\}$, where the maximum and minimum eigenvalues  are denoted by $\lambda_{max}\left\{M\right\}$ and $\lambda_{min}\left\{M\right\}$  respectively. Moreover,  by $det M$ or $|M|$  we mean the determinant  of the matrix $M$.
	
	\begin{assumption}\label{assum1}
		Let $\left\{\mathcal{F}_{k},k\geq 0\right\}$ be a non-decreasing sequence of $\sigma -$algebras such that $\phi_{k}$ is $\mathcal{F}_{k}-$measurable with a known upper bound:
		\begin{equation}\label{4}
			\sup _{k \geq 1}\left\|\phi_{k}\right\| = M<\infty, \qquad a.s.
		\end{equation}
		where $M$ may be a random variable.
	\end{assumption}
	\begin{assumption}\label{assum2}
		The true parameter $\theta$ belongs to a bounded convex set $D \subseteq \mathcal{R}^{n}$, and we denote
		\begin{equation}\label{5}
			\sup _{x \in D}\|x\| = L<\infty, \qquad a.s.
		\end{equation}
	\end{assumption}
	\begin{assumption}\label{assum3}
		The given threshold $\left\{c_{k},\mathcal{F}_{k}\right\}$ is an adapted sequence, with a known upper bound:
		\begin{equation}\label{a17}
			\sup_{k\geq 0}|c_{k}|=C <\infty,\qquad	a.s.
		\end{equation} 	
		where $C$ may be a random variable.	
	\end{assumption}
	\begin{assumption}\label{assum4}
		The noise $v_{k}$ is  integrable and $\mathcal{F}_{k}-$measurale. For any $k\geq 1$, the conditional probability density function of $v_{k}$ given $\mathcal{F}_{k-1}$, denoted by $f_{k}(\cdot)$, is known and satisfies
		\begin{equation}\label{6}
			\begin{aligned}
				\inf_{|x| \leq LM+C} \left\{f_{k}(x)\right\} > 0,\; \;\;\; k=0,1,\cdots, \qquad a.s.
			\end{aligned}
		\end{equation}
		where $L$, $M$ and $C$ are defined by $(\ref{4})$, $(\ref{5})$ and $(\ref{a17})$.
	\end{assumption}
	\begin{remark}
		It can be easily seen that if the threshold $c_{k}$ is fixed, then Assumption $\ref{assum3}$ will be satisfied automatically. Moreover, if the noise $v_{k}$ is independent with the $\sigma-$algebra $\mathcal{F}_{k-1}$, and with identically  normal distribution as assumed previously (see,e.g., \cite{guo:2013}, \cite{zhang:2019}), then the condition $(\ref{6})$ in Assumption $\ref{assum4}$ will be satisfied.
	\end{remark}
	\subsection{Recursive algorithm and adaptive predictor}
	To construct a Quasi-Newton type identification algorithm, we need to introduce a projection operator on $R^{p}$ as follows.
	\begin{definition}
		For the linear space $R^{p}~ (p\geq1)$, the weighted norm $\|\cdot \|_{Q}$ associated with a positive definite matrix $Q$ is defined as
		\begin{equation}\label{7}
			\| x\|_{Q}=x^{\tau}Qx, \quad \forall x \in \mathcal{R}^{p}.
		\end{equation}
	\end{definition}
	\begin{definition}\label{def2}
		For a given convex compact set $\Omega \in R^{p}$, and a positive definite matrix Q, the projection operator $\Pi_{Q}(\cdot)$ is defined as
		\begin{equation}\label{8}
			\Pi_{Q}(x)=\mathop{\arg\min}_{\omega \in \Omega}\|x-\omega\|_{Q}, \quad \forall x \in \mathcal{R}^{p}
		\end{equation}
	\end{definition}
	\begin{remark}
		The well-posedness of  $ \Pi_{Q}(x)$ is ensured by the positive definite property of the matrix $Q$ and the convexity of $\Omega$ \citep{cheney:2001}.
	\end{remark}
	
	Our recursive identification algorithm is a kind of Quasi-Newton algorithm, defined  as follows:
	\begin{alg}\label{alg1}
		\begin{equation}\label{9}
			\hat{\theta}_{k+1}=\Pi_{P_{k+1}^{-1}}\left\{\hat{\theta}_{k}+a_{k}\beta_{k}P_{k}\phi_{k}e_{k+1}\right\},
		\end{equation}
		\begin{equation}\label{a1}
			P_{k+1}=P_{k}-\beta_{k}^{2}a_{k}P_{k}\phi_{k}\phi_{k}^{\tau}P_{k},
		\end{equation}
		\begin{equation}\label{a2}
			e_{k+1}=s_{k+1}-1+F_{k+1}(c_{k}-\phi_{k}^{\mathrm{\tau}} \hat{\theta}_{k}),
		\end{equation}
		\begin{equation}\label{a3}
			a_{k}=\frac{1}{1+\beta_{k}^{2}\phi_{k}^{\tau}P_{k}\phi_{k}},
		\end{equation}
		\begin{equation}\label{a4}
			0<\beta_{k+1}\leq \min{\left\{\beta_{k}, \inf _{|x| \leq LM+C}f_{k+2}(x)\right\}},
		\end{equation}
	\end{alg}
	where $\hat{\theta}_{k}$ is the estimate of $\theta$ at
	time $k$; $\Pi_{P_{k}^{-1}}$ is a projection operator defined as in Definition $\ref{def2}$;   $F_{k}$ is the conditional probability distribution function of $v_{k}$ given the $\sigma-$algbra $\mathcal{F}_{k-1}$;  the initial value $\hat{\theta}_{0}$ can be chosen arbitrarily in $D$, where $D$ is given in Assumption $\ref{assum2}$; $\beta_{0} $ can be arbitrarily chosen from the interval $\left(0, \min{\left\{1, \inf _{|x| \leq LM+C}f_{1}(x)\right\}}\right)$ ; $P_{0}>0$ can also be chosen arbitrarily.
	
	Note that by $(\ref{a1})$ and the well-known matrix inversion formula (see, e.g., \cite{guo:2020}, Theorem 1.1.17), the inverse of $P_{k}$ can be recursively rewritten as
	\begin{equation}\label{10}
		P_{k+1}^{-1}=P_{k}^{-1}+\beta_{k}^{2}\phi_{k}\phi_{k}^{\tau},\;\;\;\; k=0,1,\cdots.
	\end{equation}
	
	Thus, $P_{k}^{-1}$ is positive-definite since the initial condition $P_{0}>0$, which ensures the well-posedness of the projection operator $\Pi_{P_{k}^{-1}}$ in Algorithm $\ref{alg1}$.
	
	Moreover, since both $c_{k}$ and $\phi_{k}$ are $\mathcal{F}_{k}-$measurable, we have
	\begin{equation}\label{a15}
		E(y_{k+1}\mid \mathcal{F}_{k})=\theta^{\tau}\phi_{k} +E\left(v_{k+1}\mid \mathcal{F}_{k}\right)
	\end{equation}
	which is the best prediction for $y_{k+1}$ given $\mathcal{F}_{k}$ in the mean square sense. Note that $E\left(v_{k+1}\mid \mathcal{F}_{k}\right)$ can be obtained by the known conditional probability density function $f_{k+1}\left(\cdot\right)$ in Assumption $\ref{assum4}$. Replacing the unknown parameter $\theta$ in $(\ref{a15})$ by its estimate $\hat{\theta}_{k}$, we can obtain a natural adaptive predictor of $y_{k+1}$  as follows:
	\begin{equation}\label{11}
		\hat{y}_{k+1}=\hat{\theta}_{k}^{\tau}\phi_{k} +E\left(v_{k+1}\mid \mathcal{F}_{k}\right).
	\end{equation}
	The difference between the best prediction and adaptive prediction can be regarded as regret, denoted as  $R_{k}$, i.e.,
	\begin{equation}\label{12}
		\begin{aligned}
			R_{k}=&\left[E(y_{k+1}\mid \mathcal{F}_{k})-\hat{y}_{k+1} \right]^{2}=\left(\tilde{\theta}_{k}^{\tau}\phi_{k}\right)^{2},
		\end{aligned}
	\end{equation}
		where $\tilde{\theta}_{k}=\theta-\hat{\theta}_{k}$. One may naturally expect that the regret $R_{k}$  be small in some sense, which will be useful in adaptive control. Details will be discussed in the subsequent section.
	
	Throughout the sequel, for convenience , let us introduce the following notations:
	\begin{equation}\label{cc}
		\gamma_{k}=1/\beta_{k},
	\end{equation}
	\begin{equation}\label{23}
		\omega_{k+1}=s_{k+1}-1+F_{k+1}(c_{k}-\theta^{\tau}\phi_{k}),
	\end{equation}
	\begin{equation}\label{24}
		\psi_{k}=F_{k+1}(c_{k}-\hat{\theta}_{k}^{\tau}\phi_{k})-F_{k+1}(c_{k}-\theta^{\tau}\phi_{k})).
	\end{equation}
	
	\subsection{Global convergence results}
	The following three theorems are the main results of this paper.  Under no excitation conditions, we will establish some nice asymptotic upper bounds for the parameter estimation error, the accumulated regrets of adaptive prediction, and the tracking error of adaptive control.
	\begin{theorem}\label{thm2}
		Under Assumptions $\ref{assum1}$-$\ref{assum4}$, the estimation error produced by  the estimation Algorithm $\ref{alg1}$ has the following upper bound:
		\begin{equation}\label{15}
			\begin{aligned}
				\left\|\tilde{\theta}_{n+1}\right\|^{2}=&O\left(\frac{ \log \left(\lambda_{\max}\left\{P_{n+1}^{-1}\right\}  \right)}{\lambda_{\min }\left\{P_{n+1}^{-1}\right\}}\right), \;\;\mathrm{a.s}.
			\end{aligned}
		\end{equation}
		where $\tilde{\theta}_{k}=\theta-\hat{\theta}_{k}$.
	\end{theorem}
	The detailed proof of Theorem $\ref{thm2}$ is supplied in the next section.
	
	\begin{corollaryx}\label{cor3}
		Let the conditions of Theorem $\ref{thm2}$ hold, and let the conditional probability density function $\left\{f_{k}(x)\right\}$ of the noise sequence have a uniformly positive lower bound:
					\begin{equation}\label{dd9}
				\inf_{|x|\leq LM+C, k\geq 0}\left\{ f_{k}(x)\right\} > 0,\;\;\mathrm{a.s}.
			\end{equation}
			Then
			\begin{equation}\label{dd8}
				\left\|\tilde{\theta}_{n+1}\right\|^{2}=O\left(\frac{\log n }{\lambda_{\min }\left\{P_{0}^{-1}+\sum_{i=1}^{n} \phi_{i} \phi_{i}^{\tau}\right\}}\right),\;\;\mathrm{a.s}.	
			\end{equation}		
	\end{corollaryx}

	\begin{remark}\label{re555}
		Let the noise $v_{k}$ be independent with $\sigma-$algebra $\mathcal{F}_{k-1}$, and normally distributed with zero mean and variance $\sigma_{k}^{2}$, $k\geq 1$. Then the condition $(\ref{dd9})$ will be satisfied if $\left\{\sigma_{k}^{2}\right\}$ has an upper and positive lower bound.
	\end{remark}
	\begin{remark}\label{re55}
		From  $(\ref{dd8})$ we know that if we have
		\begin{equation}\label{ccc}
			\log n=o\left(\lambda_{\min }\left\{\sum_{i=1}^{n} \phi_{i} \phi_{i}^{\tau}\right\}\right),\;\; \mathrm{a.s}.
		\end{equation}
		as $n\rightarrow \infty$, then the estimates given by Algorithm $\ref{alg1}$ will be strongly consistent, i.e., $\lim_{n\rightarrow \infty}\|\tilde{\theta}_{n}\|=0,\;a.s.$
		The condition $(\ref{ccc})$ is much weaker than the traditional persistent excitation condition, which requires that\,  $n=O\left(\lambda_{min}\left\{\sum \limits_{i=1}^{n}\phi_{i}\phi_{i}^{\tau} \right\}\right)$.  Also, the condition $(\ref{ccc})$ is equal to the Lai-Wei excitation condition $(\ref{3})$ for classical least-squares algorithm with regular output sensors.
	\end{remark}
	
	\begin{theorem}\label{thm1}
		 Consider the estimation Algorithm $\ref{alg1}$ under Assumptions $\ref{assum1}$-$\ref{assum4}$. The sample paths of the accumulated regrets will have the following upper bound:
		\begin{equation}\label{14}
			\sum_{k=0}^{n} R_{k}=O\left(\gamma_{n}^{2}\log |P_{n+1}^{-1}|\; \right),\;\;a.s.
		\end{equation}
		where $R_{k}$, $\gamma_{k}$ are defined by $(\ref{12})$ and $(\ref{cc})$, respectively.
	\end{theorem}
	The proof of Theorem $\ref{thm1}$ is given in Section $\ref{sec3}$.
	
	According to Theorem $\ref{thm1}$, one can directly deduce the following corollary.
	
	\begin{corollaryx}\label{cor2}
		Let the conditions of Theorem $\ref{thm1}$ hold, and let  $\left\{f_{k}(x)\right\}$  be the conditional probability density function of the noise sequence as defined in Assumption $\ref{assum4}$. Then we have the following two basic results  for the accumulated regret of adaptive prediction:
		\begin{itemize}
			\item If $\left\{f_{k}(x)\right\}$ has a uniformly  positive lower bound, i.e.
			\begin{equation}\label{dd1}
				\inf_{|x|\leq LM+C, k\geq 0}\left\{ f_{k}(x)\right\} > 0,\;\; \mathrm{a.s}.
			\end{equation}
			then
			\begin{equation}\label{ddd}
				\sum_{k=0}^{n} R_{k}= O(\log n),\;\; \mathrm{a.s}.	
			\end{equation}
			\item If $\left\{f_{k}(x)\right\}$ does not have a uniformly positive lower bound but satisfies
			\begin{equation}\label{dd2}
				\sqrt{\frac{\log  k}{k}} = o\;\left(\inf_{|x| \leq LM+C} \left\{ f_{k}(x)\right\}\right),\;\; \mathrm{a.s}.
			\end{equation}
			then
			\begin{equation}\label{bb}
				\sum_{k=0}^{n} R_{k} = o(n),\;\; \mathrm{a.s}.
			\end{equation}
			
		\end{itemize}
	\end{corollaryx}
	\begin{remark}\label{rem5}
		Let the noise sequence $\left\{v_{k}\right\}$ be independent and normally distributed with zero mean and variance $\left\{\sigma_{k}^{2}\right\}$. Then the condition $(\ref{dd1})$ will be satisfied if $\left\{\sigma_{k}^{2}\right\}$ has both upper and lower  positive bounds;  the conditions  $(\ref{dd2})$ will be satisfied if $\sigma_{k}^{2}\rightarrow 0$ and $\sigma_{k}^{2}\log k \rightarrow \infty$.
	\end{remark}
	\begin{remark}
		The result $(\ref{ddd})$ in Corollary $\ref{cor2}$ is similar to the result for the classical LS algorithm for linear stochastic regression models with regular sensors, where the order $O(\log n)$ for the accumulated regrets is the best possible among all adaptive predictors(see \cite{lai:1986}).
		\end{remark}

 As in the regular observation case  (see \cite{guo:1995}), an important application of Theorem $\ref{thm1}$ is in adaptive control of stochastic systems with binary-valued observations, as stated in the following theorem:
	\begin{theorem}\label{thm3}
Let the conditions of Theorem $\ref{thm1}$ hold. And let the conditional probability density function $\left\{f_{k}(x)\right\}$ satisfy $(\ref{dd1})$ and
\begin{equation}\label{uuu}
	\sup_{k}E\left[|v_{k}|^{\alpha}\mid \mathcal{F}_{k-1}\right]<\infty, \quad a.s.,
\end{equation}
for some $\alpha > 4$. If the regression vectors $\phi_{k}$ can be influenced by an input signal $u_{k}$, such that for a given bounded sequence of reference signals $\{y_{k+1}^{*}\}$, the following equation can be satisfied by choosing $u_{k}$:
\begin{equation}\label{ggg}
	\hat{\theta}_{k}^{\tau}\phi_{k}+E(v_{k+1}\mid\mathcal{F}_{k})=y_{k+1}^{*}.
\end{equation}
Then the averaged tracking error $J_{n}$, defined by
\begin{equation}
J_{n}=\frac{1}{n}\sum_{k=0}^{n-1}\left(y_{k+1}-y_{k+1}^{*}\right)^{2},
\end{equation}
will approach to its minimum value ${\frac{1}{n}}{\sum \limits_{k=1}^{n}\sigma_{k}^{2}}$ with the following best possible almost sure convergence rate:
\begin{equation}\label{ffff}
	\begin{aligned}
		\left|J_{n}-\frac{1}{n}\sum_{k=1}^{n}\sigma_{k}^{2}\right|=O\left(\sqrt{\frac{\log \log n}{n}}\right),\quad a.s.
	\end{aligned}
\end{equation}
where $\sigma_{k}^{2}=E\left\{[v_{k}-E(v_{k} \mid \mathcal{F}_{k-1})]^{2}\mid \mathcal{F}_{k-1}\right\}$.
	\end{theorem}

The detailed proof of Theorem $\ref{thm3}$ is given in the next section.

	\section{Proofs of the main results}\label{sec3}
	To prove the main results, we first introduce several lemmas.
	
	\begin{lemma}\label{lem1}	\citep{cheney:2001}.The projection operator given by Definition $\ref{def2}$  satisfies
		\begin{equation}\label{17}
			\|\Pi_{Q}(x)-\Pi_{Q}(y)\|_{Q} \leq \|x-y\|_{Q}\quad \forall x, y\in R^{p}
		\end{equation}
	\end{lemma}
	
\begin{lemma}\label{lem2}	\citep{chen:1991}.Let $\left\{\omega_{n}, \mathcal{F}_{n}\right\}$ be a martingale difference sequence and $\left\{f_{n}, \mathcal{F}_{n}\right\}$ an adapted sequence.
	If
	\begin{equation}\label{18}
		\sup_{n} E[|\omega_{n+1}|^{\alpha}\mid \mathcal{F}_{n}] < \infty \;\;a.s.
	\end{equation}
	for some $\alpha \in (0, 2]$, then as $n\rightarrow \infty$:
	\begin{equation}\label{19}
		\sum_{i=0}^{n}f_{i}\omega_{i+1} = O(s_{n}(\alpha)\log^{\frac{1}{\alpha}+\eta}(s_{n}^{\alpha}(\alpha)+e))\;a.s., \forall \eta >0,
	\end{equation}
	where
	\begin{equation}
		s_{n}(\alpha)=\left(\sum_{i=0}^{n}|f_{i}|^{\alpha}\right)^{\frac{1}{\alpha}}
	\end{equation}
\end{lemma}
	\begin{lemma}\label{lem3}	\citep{lai:1982}.
		Let \;$X_{1}, X_{2},\cdots$ be a sequence of vectors in $R^{p} (p\geq 1)$ and let $A_{n} = A_{0}+\sum_{i=1}^{n}X_{i}X_{i}^{\tau}$. Let $|A_{n}|$ denote the determinant of $A_{n}$. Assume that $A_{0}$ is nonsingular, then as $n\rightarrow \infty$
		\begin{equation}\label{20}
			\sum_{k=0}^{n}\frac{X_{k}^{\tau}A_{k}^{-1}X_{k}}{1+X_{k}^{\tau}A_{k}^{-1}X_{k}} = O(\log(|A_{n}|)).	
		\end{equation}
	\end{lemma}
	\begin{lemma}\label{lem4} \citep{guo:1995}.
		Let \;$X_{1}, X_{2},\cdots$ be any bounded sequence of vectors in $R^{p}~(p\geq 1)$.  Denote $A_{n} = A_{0}+\sum_{i=1}^{n}X_{i}X_{i}^{\tau}$ with $A_{0} > 0 $, then we have
		\begin{equation}\label{b1}
			\sum_{k=0}^{\infty}\left(X_{k}^{\tau}A_{k}^{-1}X_{k}\right)^{2}  < \infty.			
		\end{equation}
	\end{lemma}
	Finally,  the proofs of  Theorems $\ref{thm2}$-$\ref{thm3}$ will immediately follow from the following Lemma $\ref{lem6}$, which can be proven by using Lemmas \ref{lem1}-\ref{lem4}. 	

	\begin{lemma}\label{lem6}
		Let Assumptions $\ref{assum1}$-$\ref{assum4}$ be satisfied. Then the parameter estimate given by  Algorithm $\ref{alg1}$ has the following property as n $\rightarrow \infty$:
		\begin{equation}\label{22}
			\begin{aligned}
				\tilde{\theta}_{n+1}^{\tau}P_{n+1}^{-1}\tilde{\theta}_{n+1}+\beta_{n}^{2}\sum_{k=0}^{n}\left(\tilde{\theta}_{k}^{\tau}\phi_{k} \right)^{2}	= O\left(\log |P_{n+1}^{-1}|\right).
			\end{aligned}
		\end{equation}
	\end{lemma}
	{\bf Proof.}	
	By Assumptions $\ref{assum1}$-$\ref{assum4}$ and (\ref{9})-(\ref{a4}), $\psi_{k}$ is $\mathcal{F}_{k}-$measurable, and satisfies
	\begin{equation}\label{25}
		|\psi_{k}|\leq 1, k=0,1,\cdots
	\end{equation}
	Moreover, by $(\ref{a15})$ and $(\ref{23})$ 
	\begin{equation}\label{26}
		E(\omega_{k+1}\mid \mathcal{F}_{k})=0,
	\end{equation}
	which means $\left\{{\omega_{k}}, \mathcal{F}_{k}  \right\}$ is a martingale difference sequence.
	
	Following the analysis ideas of the classical least-squares for linear stochastic regression models(see e.g., \cite{moore:1978}, \cite{lai:1982}, \cite{guo:1995}), we consider the following stochastic  Lyapunov function: $$ V_{k}=\tilde{\theta}_{k}^{\tau}P_{k}^{-1}\tilde{\theta}_{k}. $$By  Lemma $\ref{lem1}$ and Algorithm $\ref{alg1}$ ,
	we know that
	\begin{equation}\label{27}
		\begin{aligned}
			&V_{k+1}=\tilde{\theta}_{k+1}^{\tau}P_{k+1}^{-1}\tilde{\theta}_{k+1} \\
			\leq& \left\{\tilde{\theta}_{k}-a_{k}P_{k}\beta_{k}\phi_{k}\left[s_{k+1}-1+F_{k+1}\left(c_{k}-\phi_{k}^{\mathrm{\tau}} \hat{\theta}_{k}\right)\right] \right\}^{\tau}P_{k+1}^{-1}\cdot \\
			&\left\{\tilde{\theta}_{k}-a_{k}P_{k}\beta_{k}\phi_{k}\left[s_{k+1}-1+F_{k+1}\left(c_{k}-\phi_{k}^{\mathrm{\tau}} \hat{\theta}_{k}\right)\right] \right\}\\
			=&\left\{\tilde{\theta}_{k}-a_{k}P_{k}\beta_{k}\phi_{k}\left[\psi_{k}+\omega_{k+1}\right]\right\}^{\tau}P_{k+1}^{-1}\cdot \\
			&\left\{\tilde{\theta}_{k}-a_{k}P_{k}\beta_{k}\phi_{k}\left[\psi_{k}+\omega_{k+1}\right]\right\}\\
			=&\tilde{\theta}_{k}^{\tau}P_{k+1}^{-1}\tilde{\theta}_{k}-2a_{k}\beta_{k}\tilde{\theta}_{k}^{\tau}P_{k+1}^{-1}P_{k}\phi_{k}\psi_{k}\\
			&+a_{k}^{2}\beta_{k}^{2}\psi_{k}^{2}\phi_{k}^{\tau}P_{k}P_{k+1}^{-1}P_{k}\phi_{k}\\
			&+2a_{k}^{2}\beta_{k}^{2}\psi_{k}\phi_{k}^{\tau}P_{k}P_{k+1}^{-1}P_{k}\phi_{k}\omega_{k+1}\\
			&-2a_{k}\beta_{k}\phi_{k}^{\tau}P_{k}P_{k+1}^{-1}\tilde{\theta}_{k}\omega_{k+1}\\
			&+a_{k}^{2}\beta_{k}^{2}\phi_{k}^{\tau}P_{k}P_{k+1}^{-1}P_{k}\phi_{k}\omega_{k+1}^{2}.
		\end{aligned}
	\end{equation}
	
	Let us now analyze the right-hand-side (RHS) of $\left(\ref{27} \right)$ term by term. From (\ref{10}), we know that
	\begin{equation}\label{xx}
	\tilde{\theta}_{k}^{\tau}P_{k+1}^{-1}\tilde{\theta}_{k} = \tilde{\theta}_{k}^{\tau}P_{k}^{-1}\tilde{\theta}_{k}+\beta_{k}^{2}(\tilde{\theta}_{k}^{\tau}\phi_{k})^{2}.
	\end{equation}
	Moreover, by (\ref{10}) again, we know that
	\begin{equation}\label{28}
		\begin{aligned}
			&a_{k}P_{k+1}^{-1}P_{k}\phi_{k}\\
			=&a_{k}\left(I+\beta_{k}^{2}\phi_{k}\phi_{k}^{\tau}P_{k}\right)\phi_{k}\\
			=&a_{k}\phi_{k}\left(1+\beta_{k}^{2}\phi_{k}^{\tau}P_{k}\phi_{k}\right)\\
			=&\phi_{k}.
		\end{aligned}
	\end{equation}
	Hence, we have
	\begin{equation}\label{29}
		\begin{aligned}
			&2a_{k}\beta_{k}\tilde{\theta}_{k}^{\tau}P_{k+1}^{-1}P_{k}\phi_{k}\psi_{k}\\
			=&2\beta_{k}\tilde{\theta}_{k}^{\tau}\phi_{k}\psi_{k} = 2\beta_{k}|\tilde{\theta}_{k}^{\tau}\phi_{k}|\cdot|\psi_{k}|\\
			\geq &2\beta_{k}^{2}(\tilde{\theta}_{k}^{\tau}\phi_{k})^{2},
		\end{aligned}
	\end{equation}
	where the second equality is since $F_{k+1}(\cdot)$ is an increasing function, and the last inequality holds by ($\ref{a4}$) and
	\begin{equation}\label{30}
		|\psi_{k}|=\left|\int_{c_{k}-\theta^{\tau}\phi_{k}}^{c_{k}-\hat{\theta}_{k}^{\tau}\phi_{k}}f_{k+1}(x)dx\right| \geq \beta_{k}|\tilde{\theta}_{k}^{\tau}\phi_{k}|.
	\end{equation}
	Similarly, by $(\ref{28})$,
	\begin{equation}\label{31}
		\begin{aligned}
			& a_{k}^{2}\beta_{k}^{2}\psi_{k}^{2}\phi_{k}^{\tau}P_{k}P_{k+1}^{-1}P_{k}\phi_{k}\\
			=&a_{k}\beta_{k}^{2}\psi_{k}^{2}\phi_{k}^{\tau}P_{k}\phi_{k}	\leq a_{k}\beta_{k}^{2}\phi_{k}^{\tau}P_{k}\phi_{k}
		\end{aligned}
	\end{equation}
	where we have used the fact that $|\psi_{k}| \leq 1$.
	
	Now, substituting $(\ref{xx})$, $(\ref{29})$ and  $(\ref{31})$ into $(\ref{27})$ we get
	\begin{equation}\label{ee1}
		\begin{aligned}
			V_{k+1}\leq&\tilde{\theta}_{k}^{\tau}P_{k}^{-1}\tilde{\theta}_{k}-\beta_{k}^{2}(\tilde{\theta}_{k}^{\tau}\phi_{k})^{2}
			+a_{k}\beta_{k}^{2}\phi_{k}^{\tau}P_{k}\phi_{k}\\
			&+2a_{k}^{2}\beta_{k}^{2}\psi_{k}\phi_{k}^{\tau}P_{k}P_{k+1}^{-1}P_{k}\phi_{k}\omega_{k+1}\\
			&-2a_{k}\beta_{k}\phi_{k}^{\tau}P_{k}P_{k+1}^{-1}\tilde{\theta}_{k}\omega_{k+1}\\
			&+a_{k}^{2}\beta_{k}^{2}\phi_{k}^{\tau}P_{k}P_{k+1}^{-1}P_{k}\phi_{k}\omega_{k+1}^{2},
		\end{aligned}
	\end{equation}

	Summing up both sides of $\left(\ref{ee1} \right)$ from 0 to $n$, we have
	\begin{equation}\label{32}
		\begin{aligned}
			V_{n+1}\leq &\tilde{\theta}_{0}^{\tau}P_{0}^{-1}\tilde{\theta}_{0}+\sum_{k=0}^{n}a_{k}\beta_{k}^{2}\phi_{k}^{\tau}P_{k}\phi_{k}\\
			&-\beta_{k}^{2}\sum_{k=0}^{n}(\tilde{\theta}_{k}^{\tau}\phi_{k})^{2}\\
			&+2\sum_{k=0}^{n}a_{k}^{2}\beta_{k}^{2}\psi_{k}\phi_{k}^{\tau}P_{k}P_{k+1}^{-1}P_{k}\phi_{k}\omega_{k+1}\\
			&-2\sum_{k=0}^{n}a_{k}\beta_{k}\phi_{k}^{\tau}P_{k}P_{k+1}^{-1}\tilde{\theta}_{k}\omega_{k+1}\\
			&+\sum_{k=0}^{n}a_{k}^{2}\beta_{k}^{2}\phi_{k}^{\tau}P_{k}P_{k+1}^{-1}P_{k}\phi_{k}\omega_{k+1}^{2}
		\end{aligned}
	\end{equation}
	We now analyze the last three  terms in $(\ref{32})$ which are related to the martingale difference sequence $\left\{\omega_{k}, \mathcal{F}_{k} \right\}$.
	
	Let  $X_{k}=\beta_{k}\phi_{k}$ in Lemma $\ref{lem3}$ and Lemma $\ref{lem4}$, we get
		\begin{equation}\label{35}
		\sum_{k=0}^{n}a_{k}\beta_{k}^{2}\phi_{k}^{\tau}P_{k}\phi_{k}=O(\log \left|P_{n+1}^{-1}\right|),
	\end{equation}
		\begin{equation}\label{fff}
		\sum_{k=0}^{n}(\beta_{k}^{2}\phi_{k}^{\tau}P_{k}\phi_{k})^{2} = O(1),	
	\end{equation}
		respectively. Moreover, since $|\omega_{k}|\leq 1$, we have
		\begin{equation}\label{final}
		\sup_{k} E[|\omega_{k+1}|^{2}\mid \mathcal{F}_{k}] < \infty, \;a.s.
		\end{equation}
		 Denote
		\begin{equation}
		\tilde{S}_{n}=\sqrt{\sum_{k=0}^{n}\left(a_{k}\beta_{k}^{2}\psi_{k}\phi_{k}^{\tau}P_{k}\phi_{k}\right)^{2}}.
		\end{equation}
		 By  $(\ref{fff})$ and the boundedness of $a_{k}$ and $\psi_{k}$, we know that $\tilde{S}_{n} = O(1)$.   Consequently, by $(\ref{28})$ and Lemma $\ref{lem2}$, we have
	\begin{equation}\label{33}
		\begin{aligned}
			&\sum_{k=0}^{n}a_{k}^{2}\beta_{k}^{2}\psi_{k}\phi_{k}^{\tau}P_{k}P_{k+1}^{-1}P_{k}\phi_{k}\omega_{k+1}\\
			=&\sum_{k=0}^{n}a_{k}\beta_{k}^{2}\psi_{k}\phi_{k}^{\tau}P_{k}\phi_{k}\omega_{k+1}\\
			=&O\left(\tilde{S_{n}} \log^{\frac{1}{2}+\eta} \left(\tilde{S}_{n}^{2}+e\right)\right) = O(1), \; \text { a.s. } \;\;\forall \eta>0.
		\end{aligned}
	\end{equation}
	Also, by Lemma $\ref{lem2}$ and $(\ref{28})$ again, we know that
		\begin{equation}\label{34}
		\begin{aligned}
			&\sum_{k=0}^{n}a_{k}\beta_{k}\phi_{k}^{\tau}P_{k}P_{k+1}^{-1}\tilde{\theta}_{k}\omega_{k+1}\\
			=&\sum_{k=0}^{n}\beta_{k}\phi_{k}^{\tau}\tilde{\theta}_{k}\omega_{k+1}\\
			=&O\left(\sum_{k=0}^{n}(\beta_{k}\tilde{\theta}_{k}^{\tau}\phi_{k})^{2} \right)^{\frac{1}{2}+\eta}\\
			=&o\left(\sum_{k=0}^{n}(\beta_{k}\tilde{\theta}_{k}^{\tau}\phi_{k})^{2} \right) + O(1) \; \text { a.s. } \;\; \forall \eta>0
		\end{aligned}
	\end{equation}
	As for the last term of right side of $(\ref{32})$,  since $\left|\omega_{k}\right|\leq 1$, we have
	\begin{equation}\label{kkk}
	  \sup _{k} E\left[\left|\omega_{k+1}^{2}-E[\omega_{k+1}^{2}\mid\mathcal{F}_{k}]\right|^{2}\mid \mathcal{F}_{k}\right] \leq 1, \;\text { a.s. },
	 \end{equation}
Denote $\Lambda_{n} =\sqrt{\sum_{k=0}^{n}\left(a_{k}\beta_{k}^{2} \phi_{k}^{\mathrm{\tau}} P_{k} \phi_{k}\right)^{2}}$, by Lemma $\ref{lem2}$ and letting $\alpha = 2$, we get
	\begin{equation}\label{36}
		\begin{aligned}
			&\sum_{k=0}^{n}a_{k}^{2}\beta_{k}^{2}\phi_{k}^{\tau}P_{k}P_{k+1}^{-1}P_{k}\phi_{k}\left\{\omega_{k+1}^{2}-E[\omega_{k+1}^{2}\mid\mathcal{F}_{k}]\right\}\\
			=&\sum_{k=0}^{n}a_{k}\beta_{k}^{2}\phi_{k}^{\tau}P_{k}\phi_{k}\left\{\omega_{k+1}^{2}-E[\omega_{k+1}^{2}\mid\mathcal{F}_{k}]\right\}\\
			=&O\left(\Lambda_{n} \log^{\frac{1}{2}+\eta} (\Lambda_{n}^{2}+e)\right)
			=O(1), \quad \text { a.s. } \forall \eta>0
		\end{aligned}
	\end{equation}
	where the last equality is from $(\ref{fff})$ and $|a_{k}|\leq 1$.
	Hence, from $(\ref{35})$ and $(\ref{final})$
	\begin{equation}\label{39}
		\begin{aligned}
			& \sum_{k=0}^{n} a_{k}\beta_{k}^{2} \phi_{k}^{\mathrm{\tau}} P_{k} \phi_{k} \omega_{k+1}^{2} \\
			\leq& \sum_{k=0}^{n}a_{k}\beta_{k}^{2}\phi_{k}^{\tau}P_{k}\phi_{k}\left(\omega_{k+1}^{2}-E[\omega_{k+1}^{2}\mid\mathcal{F}_{k}]\right)\\
			&+ \sup_{k}E[\omega_{k+1}^{2}\mid\mathcal{F}_{k}]\left( \sum_{k=0}^{n}a_{k}\beta_{k}^{2}\phi_{k}^{\tau}P_{k}\phi_{k}\right) \\
			=& O\left(\log |P_{n+1}^{-1}|) \right)\quad \text { a.s. }
		\end{aligned}
	\end{equation}
Combine $(\ref{32})$, $(\ref{33})$, $(\ref{34})$, $(\ref{39})$, we thus have
\begin{equation}
		\tilde{\theta}_{n+1}^{\tau}P_{n+1}^{-1}\tilde{\theta}_{n+1}+\sum_{k=0}^{n}\left(\beta_{k}\tilde{\theta}_{k}^{\tau}\phi_{k} \right)^{2}= O(\log |P_{n+1}^{-1}|),\quad \text { a.s. }.
\end{equation}
Note that $\left\{\beta_{k}\right\}$ is a non-increasing sequence, we finally obtain $(\ref{22})$.
	\qed

{\bf Proofs of Theorems $\ref{thm2}$ and $\ref{thm1}$.} We note that
\begin{equation}
\lambda_{min}\left\{P_{n+1}^{-1} \right\} \|\tilde{\theta}_{n+1}\|^{2} \leq \tilde{\theta}_{n+1}^{\tau}P_{n+1}^{-1}\tilde{\theta}_{n+1}.
\end{equation}
Then Theorem $\ref{thm2}$ follows from Lemma $\ref{lem6}$ immediately. Moreover, note that $\gamma_{n}=1/ \beta_{n}$, Theorem $\ref{thm1}$ also follows from Lemma $\ref{lem6}$.\qed

{\bf Proof of Theorem $\ref{thm3}$.} By the definitions of $J_{n}$, $R_{n}$ and the equation $(\ref{ggg})$, we know that
\begin{equation}\label{jjj}
		\begin{aligned}
	J_{n}=&\frac{1}{n}\sum_{k=0}^{n-1}\left[y_{k+1}-y_{k+1}^{*} \right]^{2}\\
	=&\frac{1}{n}\sum_{k=0}^{n-1}\left[y_{k+1}-\phi_{k}^{\tau}\hat{\theta}_{k}-E\left(v_{k+1}\mid \mathcal{F}_{k}\right)\right]^{2}\\
	=&\frac{1}{n}\sum_{k=0}^{n-1}R_{k}+\frac{1}{n}\sum_{k=0}^{n-1}\left[v_{k+1}-E\left(v_{k+1}\mid \mathcal{F}_{k}\right)\right]^{2}\\
	&+\frac{1}{n}\sum_{k=0}^{n-1}2\left(\phi_{k}^{\tau}\tilde{\theta}_{k}\right)\left[v_{k+1}-E\left(v_{k+1}\mid \mathcal{F}_{k}\right)\right],
		\end{aligned}
\end{equation}
We now estimate the RHS of the above equation term by term. First, by Corollary $\ref{cor2}$ we know that the first term is bounded by $O\left({\log n \over n}\right)$. For the last two terms of $(\ref{jjj})$, by Lemma $\ref{lem2}$, we have
\begin{equation}\label{hh}
		\begin{aligned}
&\sum_{k=0}^{n-1}\left(\phi_{k}^{\tau}\tilde{\theta}_{k}\right)\left[v_{k+1}-E\left(v_{k+1}\mid \mathcal{F}_{k}\right)\right]\\
=&O\left(\left\{\sum_{k=0}^{n-1}R_{k}\right\}^{\frac{1}{2}+\eta}\right) \\
=&o\left(\sum_{k=0}^{n-1}R_{k}\right) + O(1)\quad a.s.,\quad \forall \eta >0
	\end{aligned}	
	\end{equation}
Moreover, let $\xi_{k+1}=v_{k+1}-E\left(v_{k+1}\mid \mathcal{F}_{k}\right)$. From $C_{r}-$inequality and Lyapunov equality, we can get
\begin{equation}
		\begin{aligned}
	&\sup_{k}E\left[|\xi_{k+1}^{2}-E(\xi_{k+1}^{2}\mid \mathcal{F}_{k})|^{\frac{\alpha}{2}}\mid \mathcal{F}_{k} \right]\\
	\leq&2^{\frac{\alpha}{2}}\sup_{k}E\left[|\xi_{k+1}|^{\alpha}\mid \mathcal{F}_{k} \right] \\
	\leq&2^{\frac{3\alpha}{2}}\sup_{k}E\left[|v_{k+1}|^{\alpha}\mid \mathcal{F}_{k} \right]	< \infty  \quad a.s.
\end{aligned}
\end{equation}
Thus by  a refined martingale estimation theorem (see \cite{wei:1985}, \cite{guo:2020}), we have
\begin{equation}
\frac{1}{n}\sum_{k=0}^{n-1}\left[\xi_{k+1}^{2}-E\left(\xi_{k+1}^{2}\mid\mathcal{F}_{k}\right) \right]=O\left(\sqrt{\frac{\log \log n}{n}}\right), \quad a.s.	
\end{equation}
Note that $\sigma_{k+1}^{2} = E\left(\xi_{k+1}^{2}\mid\mathcal{F}_{k}\right)$, hence
\begin{equation}\label{tian}
\begin{aligned}
&\frac{1}{n}\sum_{k=0}^{n-1}\left[v_{k+1}-E\left(v_{k+1}\mid \mathcal{F}_{k}\right)\right]^{2}\\
=&\frac{1}{n}\sum_{k=0}^{n-1}\sigma_{k+1}^{2}+ \frac{1}{n}\sum_{k=0}^{n-1}\left[\xi_{k+1}^{2}-E\left(\xi_{k+1}^{2}\mid\mathcal{F}_{k}\right) \right]\\
=&\frac{1}{n}\sum_{k=0}^{n-1}\sigma_{k+1}^{2}+ O\left(\sqrt{\frac{\log \log n}{n}}\right),  \quad a.s.
\end{aligned}
\end{equation}

Combining all the results above, we can obtain $(\ref{ffff})$. The convergence rate is claimed to be the best possible because it is the same as that given by the well known iterated logarithm laws in probability theory.
\qed
	
	\section{Numerical simulation}
	{\bf Example 1.}	
	Consider the stochastic regression system
	\begin{equation}\label{a}
	 y_{k+1}=a+bu_{k}+v_{k+1},\;k=0,1,2,\cdots
	 \end{equation}
	  with binary-valued observations
	\begin{equation}\label{b}
		s_{k+1}=I_{\left[y_{k+1} \geq 0\right]}=\left\{\begin{array}{ll}1, & y_{k+1} \geq 0; \\0,&   \text { otherwise, }\end{array}\right.
	\end{equation}
    where $\left\{u_{k}\right\}$ and $\left\{v_{k}\right\}$ are the input and the noise, respectively. $\theta=[a , b]^\tau=[0.5, -0.5]^\tau \in R^{2}$ is the true parameter,  $\phi_{k}=[1,u_{k}]^{\tau}\in R^{2}$ is the corresponding regressor. The  noise sequence  $\left\{ v_{k}, k\geq 1\right\}$ is i.i.d. with standard normal distribution $\mathcal{N}(0,1)$. To estimate $\theta$ by Algorithm $\ref{alg1}$, take a convex bounded set  $$D =\left\{[x , y]^\tau : |x| \leq 2, |y| \leq 2 \right\},$$ and the initial value $\hat{\theta}_{0}=[1,-1]^\tau$, $P_{0}=I$. Moreover, the input $\left\{u_{k}, k\geq 1\right\}$ is an independent sequence  with Gaussian distribution $u_{k}\sim\sqrt[4]{\frac{1}{k}}\cdot\mathcal{N}(0,1)$, and independent of $\left\{v_{k}\right\}$.

\begin{figure}
  	\begin{center}
  		\includegraphics[width=0.52\textwidth,height=0.34\textwidth]{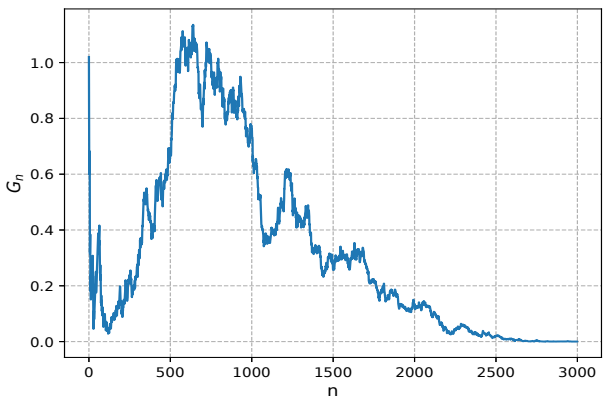}    
  		\caption{A trajectory of $G_{n}$}  
 		\label{fig1}                                 
  	\end{center}                                 
  \end{figure}

\begin{figure}
  	\begin{center}
  		\includegraphics[width=0.52\textwidth,height=0.34\textwidth]{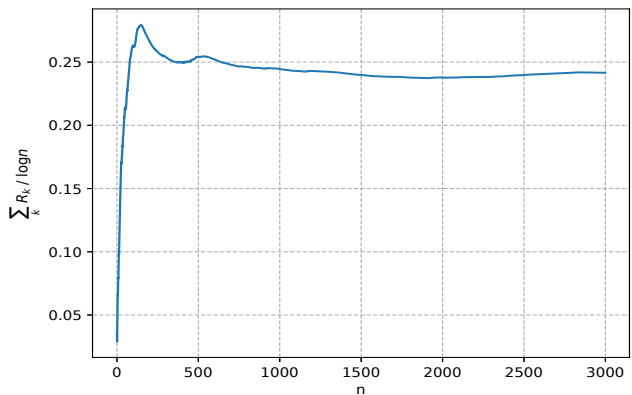}    
 		\caption{A trajectory of $\frac{1}{\log n} \sum \limits_{k=0}^{n}R_{k}$}  
		\label{fig2}                                 
 	\end{center}                                 
  \end{figure}

    In this case, we can easily verify that
    \begin{equation}\label{kkkk}
    		\begin{aligned}
    	&\lambda_{min}\left\{\sum_{k=1}^{n}\phi_{k}\phi_{k}^{\tau}\right\}\sim \sqrt{n},\\
    	&\lambda_{max}\left\{\sum_{k=1}^{n}\phi_{k}\phi_{k}^{\tau}\right\}\sim  n,
    		\end{aligned}
    \end{equation}
   as $n\rightarrow \infty$, which indicates that $\left\{\phi_{k}\right\}$ does not satisfy the PE condition. By $(\ref{kkkk})$ and $(\ref{dd8})$ in Corollary $\ref{cor3}$, the estimate error $\|\tilde{\theta}_{n}\|^{2}$ will convergent to 0 with the convergence rate $O\left(\frac{\log n}{\sqrt{n}}\right)$, which is verified by a trajectory of $G_{n}=\frac{\|\tilde{\theta}_{n}\|^{2}\sqrt{n}}{\log n}$ in Fig. \ref{fig1}. For the accumulated regrets, Fig. \ref{fig2} shows the trajectory of $\frac{1}{\log n}\cdot\sum \limits_{k=0}^{n}R_{k}$, which is bounded by $(\ref{ddd})$ in Corollary $\ref{cor2}$.

	{\bf Example 2.} Let the conditional probability density function $\left\{f_{k}(x),k\geq 1\right\}$ be defined as in Assumption $\ref{assum4}$. Then  Corollary $\ref{cor2}$ indicates that if $\inf_{|x|\leq LM+C}\left\{f_{k}(x)\right\}$ converges to 0 with proper rate, the averaged regrets of adaptive prediction will converges to 0. Let us verify this result by the model (\ref{a})-(\ref{b}) again.
		Suppose $\theta=[a,b]^\tau=[1, 1]^\tau \in R^{2}$, $\phi_{k}=[1,u_{k}]^{\tau}\in R^{2}$.  The input sequence $\left\{u_{k}, k\geq 1\right\}$ is an independent sequence with Gaussian distribution $u_{k}\sim\sqrt[4]{\frac{1}{k}}\cdot\mathcal{N}(0,1)$. The  noise sequence  $\left\{ v_{k}, k\geq 1\right\}$ is independent with  Gaussian distribution $v_{k}\sim \sqrt[4]{\frac{1}{\log k}} \cdot\mathcal{N}(0, 1)$ and independent of $\left\{u_{k}\right\}$.

\begin{figure}
	\begin{center}
		\includegraphics[width=0.52\textwidth,height=0.34\textwidth]{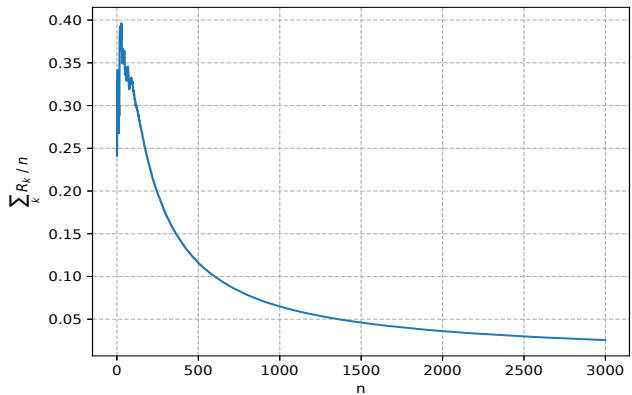}    
		\caption{A trajectory of $\frac{1}{n} \sum \limits_{k=0}^{n}R_{k}$}  
		\label{fig3}                                 
	\end{center}                                 
\end{figure}

In this case, the variance $\sigma_{k}^{2}$ of $v_{k}$ is $\sqrt{\frac{1}{\log k}}$, thus  $(\ref{dd2})$ is satisfied according to Remark $\ref{rem5}$. To construct the estimation algorithm, let the bounded parameter  set $D$  be $\left\{[x , y]^\tau : |x| \leq 3, |y| \leq 3 \right\}$, the initial value $\hat{\theta}_{0}=[-2,2]^\tau$, $P_{0}=I$. Fig. \ref{fig3} shows the convergence result $(\ref{bb})$ in  Corollary $\ref{cor2}$ for the  averaged regrets of adaptive prediction under Algorithm $\ref{alg1}$.

	{\bf Example 3.} In this example, we will give a concrete example with adaptive control to verify Theorem $\ref{thm3}$. Consider the  model (\ref{a})-(\ref{b}). Let $\theta=[a,b]^\tau=[0.5,0.8]^\tau \in R^{2}$ be the true parameter,  $\phi_{k}=[1,u_{k}]^{\tau}\in R^{2}$ be the corresponding regressor. The  noise sequence  $\left\{ v_{k}, k\geq 1\right\}$ is independent and standard normally distributed with $\sigma=1$. To give the estimate  $\hat{\theta}_{k}=\left(a_{k}, b_{k}\right)^{\tau}$ under Algorithm $\ref{alg1}$, let the bounded parameter  set $D$  be $\left\{[x , y]^\tau : |x| \leq 2, \;0.3\leq y \leq 2 \right\}$, the initial value $\hat{\theta}_{0}=[1,1]^\tau$, $P_{0}=I$.
	
\begin{figure}
	\begin{center}
		\includegraphics[width=0.52\textwidth,height=0.34\textwidth]{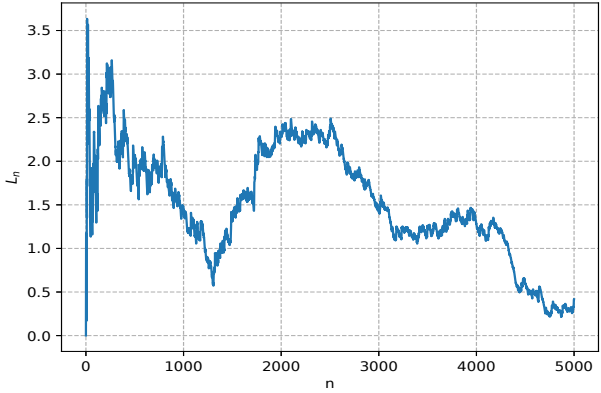}    
		\caption{A trajectory of $L_{n}$}  
		\label{fig4}                                 
	\end{center}                                 
\end{figure}

	Let the reference signals $\left\{y_{k}^{*}\right\}$ be given by $y_{k}^{*}\equiv 1$. The input $u_{k}$ is designed to ensure the output $y_{k+1}$ to track the given bounded reference signal $y_{k+1}^{*}$.  By $(\ref{ggg})$ in Theorem $\ref{thm3}$, we can design the input by solving the  following equation:
\begin{equation}
	E(y_{k+1}\mid \mathcal{F}_{k})=y_{k+1}^{*},
\end{equation}
which implies $u_{k}=(1-\hat{a}_{k})/\;\hat{b}_{k}$. From the result of the Theorem $\ref{thm3}$, the averaged tracking error $J_{n}$ will approach to $\sigma^{2}$ with the convergence rate $O\left(\sqrt{\frac{\log \log n}{n}} \right)$. Let $L_{n}=|J_{n}-\sigma^{2}|\sqrt{\frac{n}{\log \log n}}$.\; Fig.~\ref{fig4} shows the trajectory of $L_{n}$.

	\section{Concluding remarks}
	
	This paper investigated the identification and adaptation problems of  stochastic regression models with binary-valued observations, and proposed a recursive Quasi-Newton type algorithm to estimate the unknown parameters. It is shown that the estimation algorithm will converge to the true parameter almost surely under a non-persistent excitation condition. It is also  shown that the averaged  regrets of adaptive prediction converges to 0 under no excitation condition, which can be conveniently used in adaptive control to achieve asymptotic optimality of adaptive tracking without resorting to any excitation conditions. We remark that the results of this paper can be extended to various situations including any finite-valued observations with generally defined indicator functions, and that this paper may offer a necessary step for future investigations on more general adaptive control and estimation problems of stochastic systems with binary-valued observations.

\end{document}